\documentclass[journal]{IEEEtran}

\usepackage{cite,graphicx,amsmath,amssymb}
\usepackage{subfigure}
\usepackage{citesort}
\usepackage{fancyhdr}
\usepackage{mdwmath}
\usepackage{mdwtab}
\usepackage{balance}
\usepackage{xcolor}
\usepackage{bm}
 \usepackage{cases}
\usepackage{algorithm}
\usepackage{algorithmic}
\usepackage{multirow}
\usepackage{flafter}
 \usepackage{epstopdf}
\usepackage[justification=centering]{caption}

\newtheorem{remark}{\textbf{{\emph{Remark}}}}

\newtheorem{theorem}{Theorem}

\begin{document}

\title{Intelligent Reflecting Surface Enhanced Millimeter-Wave NOMA Systems}
\author{
         Jiakuo~Zuo,
        Yuanwei~Liu,~\IEEEmembership{Senior Member,~IEEE,}
        Ertugrul~Basar,~\IEEEmembership{Senior,~IEEE,}
        and Octavia A. Dobre,~\IEEEmembership{Fellow,~IEEE}
\thanks{J. Zuo is with Nanjing University of Posts and Telecommunications, Nanjing, China (email: zuojiakuo@njupt.edu.cn).}
\thanks{Y. Liu is with Queen Mary University of London, London, UK (email: yuanwei.liu@qmul.ac.uk).}
\thanks{E. Basar is with the Communications Research and Innovation Laboratory (CoreLab), Department of Electrical and Electronics Engineering, Ko\c{c} University, Sariyer 34450, Istanbul, Turkey (email: ebasar@ku.edu.tr).}
\thanks{O. A. Dobre is with the Department of Electrical and Computer Engineering, Memorial University, St. Johns, NL A1C 5S7, Canada (email: odobre@mun.ca).}
 }

\maketitle
\begin{abstract}
In this paper, a downlink  intelligent reflecting surface (IRS) enhanced millimeter-wave (mmWave) non-orthogonal multiple access (NOMA) system is considered. A joint optimization problem over active beamforming, passive beamforming and power allocation is formulated. Due to the highly coupled variables, the formulated optimization problem is non-convex. To solve this problem, an alternative optimization and successive convex approximation based iterative algorithm is proposed. Numerical results illustrate that: 1) the proposed scheme offers significant sum-rate gains, which confirms the effectiveness of introducing IRS for mmWave-NOMA systems; 2) the proposed algorithm with discrete phase shifts can achieve close performance to that of continuous phase shifts.
\end{abstract}

\vspace{-0.3cm}
\section{Introduction}
  Non-orthogonal multiple access (NOMA) is an effective technology to enhance spectrum efficiency and support massive connectivity for beyond the fifth generation wireless networks. NOMA outperforms conventional orthogonal multiple access (OMA) techniques by simultaneously sharing the communication resources between all users via the power or code domain~\cite{liu2017non}. On the other hand, the millimeter-wave (mmWave) technology has the potential to solve the bandwidth shortage problem by utilizing a great deal of spare spectrum in the high frequency range. Therefore, it is necessary to combine NOMA and mmWave to support more users and further improve the system performance~\cite{wang2019stackelberg}.

The intelligent reflecting surface (IRS) -empowered communication is a revolutionary technique to improve the network coverage, spectrum- and energy-efficiency in future wireless networks. An IRS consists of a large number of low-cost passive reflecting elements. By smartly adjusting these elements, the propagation environment can be reconfigured~\cite{wu2019towards}. For example, if the transmitter and receiver are blocked by an obstacle, an extra communication link can be created to enhance the received signal by carefully deploying IRS.

With the above benefits, IRSs have been investigated in various wireless communication systems. Specifically, the joint power control and passive beamforming optimization problem was first studied for mobile edge computing in IRS-mmWave systems  in~\cite{cao2019intelligent}. A distributed optimization algorithm was proposed to solve the joint optimization problem. In~\cite{xiu2020irs}, the joint active/passive beamforming optimization was solved by an alternating manifold optimization algorithm. In~\cite{jamali2019intelligent}, an architecture of IRS/intelligent transmitting surface assisted mmWave massive multiple-input multiple-output was designed and two efficient precoders were proposed by exploiting the sparsity of mmWave channels. The analog-digital hybrid precoding design and phase shifts optimization for IRS-mmWave was investigated in~\cite{pradhan2020hybrid} and an iterative algorithm was proposed to minimize the mean-squared-error. In~\cite{wang2019intelligent}, the joint active/passive beamforming optimization problem was studied for single- and multi-IRS assisted mmWave systems. For multiple-input single-output IRS-NOMA systems  in~\cite{zhu2019power,mu2019exploiting}, the semidefinite relaxation, second-order cone and alternating optimization were used to solve the joint active/passive beamforming optimization problem. A theoretical performance comparison between IRS-NOMA and IRS-OMA systems was provided in~\cite{zheng2020intelligent} and a low-complexity algorithm was proposed to achieve near-optimal performance. The resource allocation problem for multi-channel IRS-NOMA systems was studied in~\cite{zuo2020resource}, and an algorithm was proposed to jointly optimize the subcarrier assignment, power allocation and phase shifts. An IRS assisted uplink NOMA system was considered in~\cite{zeng2020sum} and a near-optimal solution was proposed to jointly optimize the passive beamforming and power allocation.

Inspired by the aforementioned works, it is of interest to investigate the combination of IRS, mmWave and NOMA to further enhance the wireless communications. To our best knowledge, there is no existing work on joint optimization of active/passive beamforming and power allocation for IRS assisted mmWave-NOMA systems. In this paper, we consider the IRS enhanced mmWave-NOMA systems where there are no direct links between the base station (BS) and users. With the aid of an IRS, the BS-IRS link and IRS-user links can be created to enhance the network coverage. However, the formulated optimization problem becomes non-trivial to solve, since the active/passive beamforming vectors and power allocation factors are coupled. To tackle the resultant non-convex optimization, we propose a joint optimization algorithm based on alternative optimization and successive convex approximation (SCA).

Notation: $\mathbb{C}^{M \times 1}$ denotes the set of complex vectors of size \emph{M}. $\mathbb{C}^{M \times N}$ is the set of complex matrices of size $M \times N$. diag(\textbf{x}) returns a diagonal matrix whose elements are the corresponding ones in vector \textbf{x}. ${\textbf{x}}^{T}$ and ${\textbf{x}}^{H}$ denote the transpose and conjugate transpose of vector \textbf{x}, respectively. ${\left[ \textbf{x} \right]_i}$ denotes the \emph{i}-th element of \textbf{x}, while angle(\emph{x}) denotes the phase of a complex number \emph{x}. The function real(\emph{x}) represents the real part of a complex number \emph{x}. ${\rm{E}}\left\{ \cdot \right\}$ is the expectation operator.
\section{System Model}
\begin{figure}[t!]
 \setlength{\belowcaptionskip}{-0.5cm}   
\centering
\includegraphics[width=2.3in]{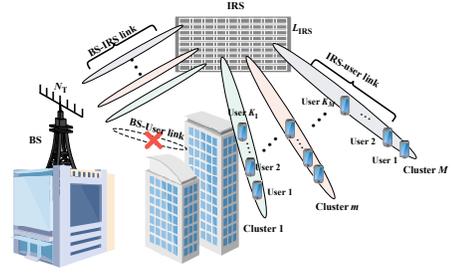}
 \caption{IRS enhanced mmWave-NOMA system.}\label{system_model}
\end{figure}

As shown in Fig.~\ref{system_model}, we consider the downlink transmission in an IRS enhanced mmWave-NOMA communication system, where the direct BS-user links are blocked and the BS communicates with \emph{K} single-antenna users with the aid of an IRS. Assume that the BS is equipped with ${N_{\rm T}}$ transmit antennas, while the IRS is equipped with ${L_{\rm IRS}}$ passive reflecting elements. The \emph{K} users are grouped into \emph{M} clusters. Let ${{\cal K}_m}$ and ${K_m}$ denote the set of the users and number of users in cluster \emph{m}, respectively. The sets of clusters and IRS passive reflecting elements are denoted by ${\cal M} = \left\{ {1,2, \cdots ,M} \right\}$ and ${\cal L_{\rm IRS}} = \left\{ {1,2, \cdots ,{L_{\rm IRS}}} \right\}$. Due to the hardware cost, the phase shifts can only be chosen from a finite set of discrete values. Specifically, the set of discrete phase shift values for each reflection element is given by: ${\theta _l} \in \Omega  \buildrel \Delta \over = \left\{ {0,\frac{{2\pi }}{{{2^B}}}, \cdots ,\frac{{2\pi \left( {{2^B} - 1} \right)}}{{{2^B}}}} \right\}$, where \emph{B} is the resolution bits of discrete phase shifts.
\vspace{-0.3cm}
\subsection{Signal Model}
For notation simplicity, we denote the \emph{k}-th user in cluster \emph{m} as user ${\cal U}\left( {m,k} \right)$. Let ${x_{m,k}}$ and ${p_{m,k}}$ be the transmitted signal and power allocation factor for ${\cal U}\left( {m,k} \right)$, where ${\rm{E}}\left\{ {{{\left| {{x_{m,k}}} \right|}^2}} \right\} = 1$. The signal received at ${\cal U}\left( {m,k} \right)$ is:
\begin{align}\label{received signal}
\begin{array}{l}
{y_{m,k}} = \underbrace {{{\textbf{h}}_{m,k}}{{\textbf{w}}_m}\sqrt {{p_{m,k}}} {x_{m,k}}}_{\rm desired~signal} + \underbrace {{{\textbf{h}}_{m,k}}{{\textbf{w}}_m}\sum\limits_{i \in \left\{ {{\mathcal{K}_m}/k} \right\}} {\sqrt {{p_{m,i}}} {x_{m,i}}} }_{\rm intra-cluster~interference}\\
 + \underbrace {{{\textbf{h}}_{m,k}}\sum\nolimits_{n \in \left\{ {\mathcal{M}\backslash m} \right\}} {{{\textbf{w}}_n}\sum\nolimits_{j \in {K_n}} {\sqrt {{p_{n,j}}} {x_{n,j}}} } }_{ \rm inter-cluster~interference} + \underbrace {{n_{m,k}}}_{\rm noise},
\end{array}
\end{align}
where $\textbf{w}_m$ is the beamforming vector for the $m$-th cluster, ${\textbf{h}_{m,k}} = \textbf{g}_{m,k}^H\boldsymbol{\Theta}\textbf{F }$ is the end-to-end channel gain for ${\cal U}\left( {m,k} \right)$, $\textbf{F} \in {{\mathbb{C}}^{{L_{\rm IRS}} \times {N_{\rm T}}}}$ is the channel matrix from the BS to the IRS,  $\boldsymbol{\Theta}  = {\rm{diag}}\left\{ {{e^{\jmath{\theta _1}}},{e^{\jmath{\theta _2}}}, \cdots ,{e^{\jmath{\theta _{{L_{\rm IRS}}}}}}} \right\}$ is the diagonal phase shifts matrix, ${\textbf{g}_{m,k}} \in {{\mathbb{C}}^{{L_{\rm IRS}} \times 1}}$ is the channel vector from the IRS to ${\cal U}\left( {m,k} \right)$, and ${{n_{m,k}}\thicksim \mathcal{{CN}}\left( {0,{\sigma ^2}} \right)}$ is the additive white Gaussian noise with zero mean and variance ${{\sigma ^2}}$.
\vspace{-0.2cm}
\subsection{Channel Model}
 We adopt the 3D Saleh-Valenzuela channel model~\cite{cao2019intelligent,pradhan2020hybrid} for the BS-IRS link and the IRS-User links. Let ${L_{\rm IRS}} = {L_{\rm x}} \times {L_{\rm z}}$, with ${L_{\rm x}}$ and ${L_{\rm z}}$ as the number of passive reflecting elements of the IRS on horizontal and vertical, respectively.  Assume that there are ${N_{\rm BI}}$ propagation paths for the BS-IRS link, $\theta _{{\rm BI},n}^{\rm AoA}$ ($\phi _{{\rm BI},n}^{\rm AoA}$) and $\theta _{{\rm BI},n}^{\rm AoD}$ are the azimuth (elevation) angle-of-arrival (AoA) and angle-of-departure (AoD) of the \emph{n}-th propagation path, respectively, $\theta _{{\rm BI},n}^{\rm AoA},\phi _{{\rm BI},n}^{\rm AoA},\theta _{{\rm BI},n}^{\rm AoD} \in \left[ { - \frac{\pi }{2}, \frac{\pi }{2}} \right]$. The channel matrix \textbf{F} is modeled as
 \begin{equation}\label{channel F}
 {\textbf{F}} = \sqrt {\frac{{{N_{\rm{T}}}{L_{{\rm{IRS}}}}}}{{{N_{{\rm{BI}}}}}}} \sum\limits_{n = 1}^{{N_{{\rm{BI}}}}} {\mathfrak{p}_n^{{\rm{BI}}}{\mathfrak{a}_n}\left( {\theta _{{\rm{BI}},n}^{{\rm{AoA}}},\phi _{{\rm{BI}},n}^{{\rm{AoA}}}} \right){{\mathfrak{b}_n^H}}\left( {\theta _{{\rm{BI}},n}^{{\rm{AoD}}}} \right)} ,
\end{equation}
where $\mathfrak{p}_n^{\rm BI}$ is the complex channel gain, $\mathfrak{a}_n\left( {\theta _{{\rm BI},n}^{\rm AoA},\phi _{{\rm BI},n}^{\rm AoA}} \right)$ and $\mathfrak{b}_n\left( {\theta _{{\rm BI},n}^{\rm AoD}} \right)$ are the array response vectors of the \emph{n}-th propagation path at the BS and IRS, which are respectively defined as
 \begin{equation}\label{aph_BI_R}
 {\mathfrak{a}_n}\left( {{\theta _n},{\phi _n}} \right) =\frac{1}{{\sqrt {{L_{{\rm{IRS}}}}} }}{\left[ {1 \cdots {e^{\jmath \mathchar'26\mkern-10mu\lambda \left( {{l_{\rm{x}}}\sin \left( {{\theta _n}} \right)\sin \left( {{\phi _n}} \right) + {l_{\rm{z}}}\cos \left( {{\phi _n}} \right)} \right)}} \cdots } \right]^T},
\end{equation}
 \begin{equation}\label{aph_BI_T}
{\mathfrak{b}_n}\left( {{\theta _n}} \right) = \frac{1}{{\sqrt {{N_T}} }}{\left[ {1\;{e^{\jmath \mathchar'26\mkern-10mu\lambda \sin \left( {{\theta _n}} \right)}}\; \cdots \;{e^{\jmath \mathchar'26\mkern-10mu\lambda \left( {{N_{\rm{T}}} - 1} \right)\sin \left( {{\theta _n}} \right)}}} \right]^T},
\end{equation}
where $\mathchar'26\mkern-10mu\lambda  = {{2\pi {d_{\rm AS}}} \mathord{\left/
 {\vphantom {{2\pi {d_{AS}}} {{\lambda _{{\rm{L}}}}}}} \right.
 \kern-\nulldelimiterspace} {{\lambda _{{\rm{L}}}}}}$, ${d_{\rm AS}}$ is the antenna spacing, ${{\lambda _{\rm L}}}$ is the carrier wavelength, $0 \le {l_{\rm x}} \le {L_{\rm x}} - 1$ and $0 \le {l_{\rm z}} \le {L_{\rm z}} - 1$.

Similarly, the channel vector ${\textbf{g}_{m,k}}$ is modeled as
 \begin{equation}\label{channel g}
 {\textbf{g}_{m,k}} = \sqrt {\frac{{{L_{\rm IRS}}}}{{{N_{\rm IU}}}}} \sum\limits_{n = 1}^{{N_{\rm IU}}} {\mathfrak{p}_{m,k,n}^{\rm IU}\boldsymbol{\mathfrak{a}}_n\left( {\theta _{{\rm IU},m,k,n}^{\rm AoD},\phi _{{\rm IU},m,k,n}^{\rm AoD}} \right)},
\end{equation}
where ${N_{\rm IU}}$ is the number of propagation paths, $\mathfrak{p}_{m,k,n}^{\rm IU} $ is the complex channel gain, $\theta _{{\rm IU},m,k,n}^{\rm AoD}$ and $\phi _{{\rm IU},m,k,n}^{\rm AoD}$ are the azimuth and elevation AoD, respectively, and ${\boldsymbol{\mathfrak{a}}_n}\left( {\theta _{{\rm IU},m,k,n}^{\rm AoD},\phi _{{\rm IU},m,k,n}^{\rm AoD}} \right)$ is the array response vector at the IRS.
In ~\eqref{channel F} and~\eqref{channel g}, \emph{n}=1 represents the line-of-sight (LoS) path and $n>1$ denotes the  non-line-of-sight (NLoS).
\vspace{-0.3cm}
\subsection{Decoding Order and Achievable Rate of Users}
In the traditional single-input single-output NOMA networks, the optimal decoding order is determined by the channel gains. However, this decoding order method cannot be used directly in IRS enhanced mmWave-NOMA systems. This is because the end-to-end channels can be modified by the IRS and the inter-cluster interference can also affect the decoding order. The optimal decoding order can be any one of the ${K_m}!$ different orders in each cluster. Therefore, an exhaustive search method is needed to find the optimal decoding order. Let ${\mathfrak{D}_m}\left( k \right)$ denote the decoding order of ${\cal U}\left( {m,k} \right)$. ${\mathfrak{D}_m}\left( k \right) = d$ means that ${\cal U}\left( {m,k} \right)$ is the \emph{d}-th signal to be decoded. Assume that the successive interference cancellation (SIC) is carried out according to the ascending order of the end-to-end channel gains.

Without loss of generality, let ${\mathfrak{D}_m}\left( k \right) = k$. For any two users ${\cal U}\left( {m,j} \right)$ and ${\cal U}\left( {m,k} \right)$ with the decoding order ${\mathfrak{D}_m}\left( j \right) > {\mathfrak{D}_m}\left( k \right)$, the signal-to-interference-plus-noise ratio (SINR) of ${\cal U}\left( {m,j} \right)$ to decode ${\cal U}\left( {m,k} \right)$ is defined as
{
 \setlength\belowdisplayskip{-2pt}
 \begin{align}\label{SINR}
 \gamma _{j \to k}^m = \frac{{{{\left| {{{{\textbf{h}}}_{m,j}}{\textbf{w}_m}} \right|}^2}{p_{m,k}}}}{{{{\left| {{{\textbf{h}}_{m,j}}{{\textbf{w}}_m}} \right|}^2}{P_{m,k}} + {\sum _{n \in \left\{ {\mathcal{M}\backslash m} \right\}}}{{\left| {{{\textbf{h}}_{m,j}}{{\textbf{w}}_n}} \right|}^2} + {\sigma ^2}}},
\end{align}
}
where ${P_{m,k}} = \sum\nolimits_{i = k + 1}^{{K_m}} {{p_{m,i}}}$. The corresponding decoding rate is $R_{j \to k}^m = {\log _2}\left( {1 + \gamma _{j \to k}^m} \right)$.

To guarantee that SIC can be performed successfully at ${\cal U}\left( {m,j} \right)$, the SIC decoding condition $R_{j \to k}^m \ge R_{m,k}^{\min }$ should be satisfied~\cite{jiang2018mimo}, with $R_{m,k}^{\min }$ as the actual information rate from BS to ${\cal U}\left( {m,k} \right)$, $\forall k \in \left\{ {1,2, \cdots ,j} \right\}$. Based on these conditions, the achievable rate of ${\cal U}\left( {m,k} \right)$ is given by~\cite{zhu2019power}
{
 \setlength\belowdisplayskip{-1pt}
 \begin{small}
\begin{numcases}{R_{m,k}=}
\mathop {{\rm{min}}}\limits_{j \in \left\{ {k, \cdots ,{K_m}} \right\}} {\log _2}\left( {1 + \gamma _{j \to k}^m} \right), {k \in \left\{ {{{\cal K}_m}\backslash {K_{m}}} \right\}} \\
 {{{\log }_2}\left( {1 + \frac{{{{\left| {{{\textbf{h}}_{m,{K_m}}}{{\textbf{w}}_m}} \right|}^2}{p_{m,{K_m}}}}}{{\sum\limits_{n \in \left\{ {\mathcal{M}\backslash m} \right\}} {{{\left| {{{\textbf{h}}_{m,{K_m}}}{{\textbf{w}}_n}} \right|}^2}}  + {\sigma ^2}}}} \right)}, k{\rm{ = }}{K_{m}}
\end{numcases}
 \end{small}
 }
  \vspace{-0.3cm}
 \section{Problem Formulation And Proposed Optimization Methods}
According to the SIC decoding condition, the quality-of-service (QoS) requirement of ${\cal U}\left( {m,k} \right)$ is given as~\cite{jiang2018mimo}:
 \begin{equation}\label{SIC condition}
\mathop {{\rm{min}}}\limits_{j \in \left\{ {k, \cdots ,{K_m}} \right\}} {\log _2}\left( {1 + \gamma _{j \to k}^m} \right) \ge R_{m,k}^{\min },
 \end{equation}
 where $k \in \left\{ {{\mathcal{K}_m}/{K_m}} \right\}, m \in {\cal M}$.

 Our goal is to maximize the sum rate of the ${K_m}$-th user, i.e., ${\cal U}\left( {m,K_m} \right)$, in each cluster under the QoS requirements of remaining users and the transmit power budget constraint. For a given decoding order, the joint active beamforming, passive beamforming and power allocation optimization problem is formulated as
\begin{subequations}\label{op}
\begin{align}
  &\mathop {\max }\limits_{\theta_l ,{\textbf{w}_m},{p_{m,k}}} \sum\nolimits_{m \in {\cal M}} {{R_{m,{K_m}}}} ,   \\
  &{s.t.} \  \sum\nolimits_{m \in {\cal M}} {\left\| {{\textbf{w}_m}} \right\|_2^2}  \le {P_{\max }}, \label{op c}\\
  &   \ \ \ \ \  \sum\nolimits_{k \in {{\cal K}_m}} {{p_{m,k}}}  = 1, m \in {\cal M}, \label{op d}\\
  &   \ \ \ \ \  {\theta _l} \in \Omega,l \in {{\cal L}_{\rm IRS}}, \label{op e} \\
  &   \ \ \ \ \  ~\eqref{SIC condition},\label{op b}
\end{align}
\end{subequations}
where ${P_{\max }}$ is the maximum transmit power.

Problem~\eqref{op} is a mixed integer non-convex optimization problem. To make it tractable, we first decouple it into three sub-problems, i.e., active beamforming optimization, passive beamforming optimization and power allocation optimization. Then, we solve them alternatively. The three subproblems are formulated as follows:
\begin{subequations}\label{ABOP}
\begin{align}
  &\mathop {\max }\limits_{{\textbf{w}_m}} \sum\nolimits_{m \in {\cal M}} {{R_{m,{K_m}}}},   \\
  &{s.t.} \ ~\eqref{op c},~\eqref{op b}. \label{ABOP b}
\end{align}
\end{subequations}
\vspace{-0.8cm}
\begin{subequations}\label{PBOP}
\begin{align}
  &\mathop {\max }\limits_{\theta_l}  \sum\nolimits_{m \in {\cal M}} {{R_{m,{K_m}}}},   \\
  &{s.t.} \ ~\eqref{op e},~\eqref{op b}. \label{PBOP b}
\end{align}
\end{subequations}
\vspace{-0.8cm}
\begin{subequations}\label{POP}
\begin{align}
  &\mathop {\max }\limits_{{p_{m,k}}} \sum\nolimits_{m \in {\cal M}} {{R_{m,{K_m}}}},   \\
  &{s.t.} \ ~\eqref{op d},~\eqref{op b}. \label{POP b}
\end{align}
\end{subequations}
\vspace{-0.9cm}
\subsection{Active Beamforming Optimization}
We first define ${\textbf{z}_{m,j}} = \textbf{g}_{m,j}^H\boldsymbol{\Theta} \textbf{F}$, ${\textbf{Z}_{m,j}}{\rm{ = }}\textbf{z}_{m,j}^H{\textbf{z}_{m,j}}$ and ${\textbf{W}_m} = {\textbf{w}_m}\textbf{w}_m^H$. Then, by introducing and substituting the auxiliary variables $\left\{ {{\chi _{m,k}}} \right\}$ into the objective function of problem~\eqref{ABOP}, the active beamforming optimization problem can be equivalently expressed as
\begin{subequations}\label{AB-SCA1}
\begin{align}
  &\mathop {\max }\limits_{{\textbf{w}_m},{\chi _{m,k}}} \sum\nolimits_{m \in \mathcal{M}} {{\rm log}{_2}\left( {1 + {\chi _{m,{K_m}}}} \right)} ,   \\
  &{s.t.} \   {\left| {{\textbf{z}_{m,j}}{\textbf{w}_m}} \right|^2}{\Delta _{p_{m,k}}} \ge r_{m,k}^{\min }{I_{m,j}}\left( {{{\textbf{w}}_n}} \right), \label{AB-SCA1-1}  \\
  &   \ \ \ \ \  {{{{\left| {{\textbf{z}_{m,{K_m}}}{\textbf{w}_m}} \right|}^2}{p_{m,{K_m}}}} \mathord{\left/
 {\vphantom {{{{\left| {{\textbf{z}_{m,{K_m}}}{\textbf{w}_m}} \right|}^2}{p_{m,{K_m}}}} {{I_{m,{K_m}}}\left( {{{\textbf{w}}_n}} \right)}}} \right. \kern-\nulldelimiterspace} {{I_{m,{K_m}}}\left( {{{\textbf{w}}_n}} \right)}} \ge {\chi _{m,{K_m}}}, \label{AB-SCA1-2} \\
  &   \ \ \ \ \  \sum\nolimits_{m \in \mathcal{M}} {\left\| {{{\textbf{w}}_m}} \right\|_2^2}  \le {P_{\max }}, \label{AB-SCA1-3}
\end{align}
\end{subequations}
where ${I_{m,j}}\left( {{{\textbf{w}}_n}} \right) = \sum\nolimits_{n \in \left\{ {\mathcal{M}/m} \right\}} {{{\left| {{\textbf{z}_{m,j}}{\textbf{w}_n}} \right|}^2}}  + {\sigma}^2 $, $r_{m,k}^{\min }{\rm{ = }}{2^{R_{m,k}^{\min }}}- 1$, ${\Delta _{{p_{m,k}}}} = {p_{m,k}} - r_{m,k}^{\min }{P_{m,k}}$,
$k \in \left\{ {{\mathcal{K}_m}/{K_m}} \right\}$, $j \in \left\{ {k,k+1, \cdots ,{K_m}} \right\}$, $m \in {\cal M}$.

The constraints~\eqref{AB-SCA1-1} and~\eqref{AB-SCA1-2} are non-convex. We first approximate constraint~\eqref{AB-SCA1-1} by the SCA method. According to the first-order Taylor series (FTS), the non-convex term in the left side of~\eqref{AB-SCA1-1} can be approximated at point ${{{\overline {\textbf{w}} }_m}}$ as
\begin{align}\label{cvx1}
 \begin{array}{l}
{\left| {{\textbf{z}_{m,j}}{\textbf{w}_m}} \right|^2} \ge 2{\rm real}\left( {\overline {\textbf{w}} _m^H{\textbf{Z}_{m,j}}{\textbf{w}_m}} \right) - {\left| {{\textbf{z}_{m,j}}{{\overline {\textbf{w}} }_m}} \right|^2}\\
 \ \ \ \ \ \ \ \ \ \ \ \ \  =\varphi_{m,j}^{{\rm{Taylor}}}\left( {{\textbf{w}_m}} \right).
\end{array}
\end{align}

For the constraint~\eqref{AB-SCA1-2}, we introduce the new variables $\left\{ {{\eta _{m,{K_m}}}} \right\}$ and decompose it into the following two inequalities
\begin{numcases}{}
  {{{{\left| {{\textbf{z}_{m,{K_m}}}{\textbf{w}_m}} \right|}^2}{p_{m,{K_m}}}} \mathord{\left/
 {\vphantom {{{{\left| {{\textbf{z}_{m,{K_m}}}{\textbf{w}_m}} \right|}^2}{p_{m,{K_m}}}} {{\eta _{m,{K_m}}}}}} \right.
 \kern-\nulldelimiterspace} {{\eta _{m,{K_m}}}}} \ge {\chi _{m,{K_m}}}, \label{equivalent 1} \\
 {\eta _{m,{K_m}}} \ge {I_{m,j}}\left( {{{\textbf{w}}_n}} \right). \label{equivalent  2}
\end{numcases}

The left side of~\eqref{equivalent 1} is a quadratic-over-affine function, which is jointly convex over the involved variables
${{\textbf{w}_m}}$ and ${{\eta _{m,{K_m}}}}$. By using the FTS approximation around ${{{\overline {\textbf{w}} }_m}}$ and ${{\eta _{m,{K_m}}}}$, we have
\begin{equation}\label{cvx2}
 \begin{array}{l}
\frac{{{{\left| {{\textbf{z}_{m,{K_m}}}{\textbf{w}_m}} \right|}^2}}}{{{\eta _{m,{K_m}}}}} \ge  \frac{{2 {\rm real}\left( {\overline {\textbf{w}} _m^H{\textbf{Z}_{m,{K_m}}}{\textbf{w}_m}} \right)}}{{{{\overline \eta  }_{m,{K_m}}}}} - {\left( {\frac{{\left| {{\textbf{z}_{m,{K_m}}}{{\overline {\textbf{w}} }_m}} \right|}}{{{{\overline \eta  }_{m,{K_m}}}}}} \right)^2}{\eta _{m,{K_m}}}\\
  \ \ \ \ \ \ \ \ \ \ \ \ \ \ = \varphi_{m,{K_m}}^{\rm{Taylor}}\left( {{\textbf{w}_m},{\eta _{m,{K_m}}}} \right).
\end{array}
\end{equation}

 From the above discussions, the active beamforming optimization problem is approximated as
\begin{subequations}\label{AB-SCA2}
\begin{align}
  &\mathop {\max }\limits_{{\textbf{w}_m},{\chi _{m,k}},{\eta _{m,{K_m}}}} \sum\nolimits_{m \in \mathcal{M}} {{\rm log}{_2}\left( {1 + {\chi _{m,{K_m}}}} \right)} ,   \\
  &{s.t.} \  \varphi_{m,j}^{{\rm{Taylor}}}\left( {{\textbf{w}_m}} \right){\Delta _{{p_{m,k}}}} \ge r_{m,k}^{\min }{I_{m,j}}\left( {{\textbf{w}_n}} \right), \label{AB-SCA2-a}  \\
  &   \ \ \ \ \  \varphi _{m,{K_m}}^{\rm{Taylor}}\left( {{\textbf{w}_m},{\eta _{m,{K_m}}}} \right){p_{m,{K_m}}} \ge {\chi _{m,{K_m}}}, \label{AB-SCA2-b} \\
  &   \ \ \ \ \  ~\eqref{AB-SCA1-3},~\eqref{equivalent  2}. \label{AB-SCA1-c}
\end{align}
\end{subequations}

It is noted that problem~\eqref{AB-SCA2} is a convex problem, which can be efficiently solved via standard convex problem solvers such as CVX~\cite{grant2014cvx}. Due to the approximations in constraints~\eqref{AB-SCA1-1} and ~\eqref{AB-SCA1-2}, problem~\eqref{AB-SCA2} is a lower bound approximation of the active beamforming problem~\eqref{ABOP}.
\vspace{-0.4cm}
\subsection{Passive Beamforming Optimization}
Before solving problem~\eqref{PBOP}, we first relax the discrete values of ${\theta _l}$ into continuous values, i.e., ${\theta _l} \in \left[ {0,2\pi } \right]$ and define the passive beamforming vector as $\textbf{v} = {\left[ {{\lambda _1}{e^{j{\theta _1}}}{\lambda _2}{e^{j{\theta _2}}} \cdots {\lambda _{{L_{\rm IRS}}}}{e^{j{\theta _{{L_{\rm IRS}}}}}}} \right]^T}$, ${\lambda _l} \in \left[ {0,1} \right]$. With the auxiliary variables $\left\{ {{\chi _{m,k}}} \right\}$, the relaxed formulation of problem~\eqref{PBOP} is formulated as
\begin{subequations}\label{equivalent PB}
\begin{align}
 &\mathop {\max }\limits_{{\textbf{v}},{\chi _{m,k}}} \sum\nolimits_{m \in \mathcal{M}} {{\rm log}{_2}\left( {1 + {\chi _{m,{K_m}}}} \right)},   \\
 &{s.t.} \  {\left| {{{{\textbf{z}}}_{m,m,j}}{\textbf{v}}} \right|^2}{\Delta _{{p_{m,k}}}} \ge r_{m,k}^{\min }{I_{m,j}}\left( {\textbf{v}} \right), \label{equivalent PB b}  \\
 &   \ \ \ \ \  {{{{\left| {{{\textbf{z}}_{m,m,{K_m}}}{\textbf{v}}} \right|}^2}{p_{m,{K_m}}}} \mathord{\left/
 {\vphantom {{{{\left| {{{\textbf{z}}_{m,m,{K_m}}}{\textbf{v}}} \right|}^2}{p_{m,{K_m}}}} {{I_{m,{K_m}}}\left( {\textbf{v}} \right)}}} \right.
 \kern-\nulldelimiterspace} {{I_{m,{K_m}}}\left( {\textbf{v}} \right)}} \ge {\chi _{m,{K_m}}}, \label{equivalent PB c}  \\
 &   \ \ \ \ \  \left| {\left[ {\textbf{v}} \right]_l} \right| \le 1, \label{equivalent PB d}
\end{align}
\end{subequations}
where ${\textbf{z}_{n,m,j}} = \textbf{g}_{m,j}^H {\rm diag}\left\{ {\textbf{F}{\textbf{w}_n}} \right\}$, ${I_{m,j}}\left( {\textbf{v}} \right) = \sum\nolimits_{n \in \left\{ {{\cal M}/m} \right\}} {{{\left| {{{\textbf{z}}_{n,m,j}}{\textbf{v}}} \right|}^2}}  + {\sigma ^2}$, $k \in \left\{ {{{\cal K}_m}/{K_m}} \right\}$, $j \in \left\{ {k,k + 1, \cdots ,{K_m}} \right\}$, $n \in \left\{ {{\cal M}/m} \right\}, m \in {\cal M}$.

Problem~\eqref{equivalent PB} is still non-convex because of the non-convexity of constraints~\eqref{equivalent PB b} and~\eqref{equivalent PB c}. We first split constraint~\eqref{equivalent PB c} into the following constraints
\begin{numcases}{}
  {{{{\left| {{{\textbf{z}}_{m,m,{K_m}}}{\textbf{v}}} \right|}^2}{p_{m,{K_m}}}} \mathord{\left/
 {\vphantom {{{{\left| {{{\textbf{z}}_{m,m,{K_m}}}{\textbf{v}}} \right|}^2}{p_{m,{K_m}}}} {{\mu _{m,{K_m}}}}}} \right.
 \kern-\nulldelimiterspace} {{\mu _{m,{K_m}}}}} \ge {\chi _{m,{K_m}}},\label{obj constraint 1}\\
  {\mu _{m,{K_m}}} \ge {I_{m,{K_m}}}\left( {\textbf{v}} \right), \label{obj constraint 2}
\end{numcases}
where $\left\{ {{\mu _{m,{K_m}}}} \right\}$ are the newly introduced variables.

Similarly, based on the FTS approximation, the inequalities~\eqref{equivalent PB b} and~\eqref{obj constraint 1} can be approximated around $\overline {\textbf{{v}}} $ and ${\overline \mu  _{m,{K_m}}}$, respectively, as:
\begin{equation}\label{SCA1}
 \left( {2\Re _{m,j}^{\left( t \right)}\left( {\textbf{v}} \right) - {{\left| {{{\textbf{z}}_{m,m,j}}{{\textbf{v}}^{\left( t \right)}}} \right|}^2}} \right){\Delta _{p_{m,k}}} \ge r_{m,k}^{\min }{I_{m,j}}\left( {\textbf{v}} \right),
\end{equation}
\begin{equation}\label{SCA2}
 \frac{{2{\Re _{m,{K_m}}}\left( {\textbf{v}} \right)}}{{{{\overline \mu  }_{m,{K_m}}}}} - {\left( {\frac{{\left| {{{\textbf{z}}_{m,m,{K_m}}}\overline {\textbf{v}} } \right|}}{{{{\overline \mu  }_{m,{K_m}}}}}} \right)^2}{\mu _{m,{K_m}}} \ge \frac{{{\chi _{m,{K_m}}}}}{{{p_{m,{K_m}}}}},
\end{equation}
where ${\Re _{m,j}}\left( {\bf{v}} \right) = {\rm{real}}\left( {{{\overline {\textbf{v}} }^H}{{\textbf{Z}}_{m,m,j}}{\textbf{v}}} \right)$ and ${\textbf{Z}_{m,m,j}} = {\textbf{z}}_{m,m,j}^H{{\textbf{z}}_{m,m,j}}$.

In summary, we have arrived at a convex approximation of the passive beamforming problem~\eqref{equivalent PB}, which is given by:
\begin{subequations}\label{equivalent PB iterative}
\begin{align}
 &\mathop {\max }\limits_{{\textbf{v}},{\chi _{m,k}},{{\mu _{m,{K_m}}}}} \sum\nolimits_{m \in \mathcal{M}} {{\rm log}{_2}\left( {1 + {\chi _{m,{K_m}}}} \right)},   \\
 &{s.t.} \  ~\eqref{equivalent PB d}, ~\eqref{obj constraint 2},~\eqref{SCA1}, ~\eqref{SCA2}.
\end{align}
\end{subequations}

Similarly, due to the approximations of constraints~\eqref{equivalent PB b} and~\eqref{equivalent PB c}, the objective value obtained form problem~\eqref{equivalent PB iterative} serves as a lower bound for the passive beamforming problem~\eqref{PBOP}.

With the solution \textbf{v} of problem~\eqref{equivalent PB iterative}, the discrete phase shifts $\left\{ {{\theta _l}} \right\}$ can be calculated via
\begin{align}\label{discrete theta}
 {\theta _l} = \arg \mathop {\min }\limits_{\theta  \in \Omega } \left| {\theta  - {\rm{\rm angle}}\left( {{{\left[ {\bf{v}} \right]}_l}} \right)} \right|.
\end{align}
\begin{remark}\label{remark:discrete}
Due to the quantization error, the discrete phase shifts $\left\{ {{\theta _l}} \right\}$ may not be a local optimal solution. However, with larger resolution bits, the quantization error becomes smaller and the discrete phase shifts can achieve the same performance as the continuous phase shifts.
To guarantee that the proposed algorithm converges, we only update $\left\{ {{\theta _l}} \right\}$  when the objective value of problem~\eqref{PBOP} is non-decreasing.
\end{remark}
\vspace{-0.3cm}
\subsection{Power Allocation Optimization}
The constraint~\eqref{op b} in problem~\eqref{POP} can be rewritten as a convex formulation
\begin{align}\label{CVX in POP}
 \left( {{p_{m,k}} - r_{m,k}^{\min }\sum\nolimits_{i = k + 1}^{{K_m}} {{p_{m,i}}} } \right){\left| {{\textbf{h}_{m,j}}{\textbf{w}_m}} \right|^2} \ge r_{m,k}^{\min }{I_{m,j}},
\end{align}
where  ${I_{m,j}} = \left( {{\sum _{n \in \left\{ {\mathcal{M}\backslash m} \right\}}}{{\left| {{{\textbf{h}}_{m,j}}{{\textbf{w}}_n}} \right|}^2} + {\sigma ^2}} \right)$, $j \in \left\{ {k,k + 1, \cdots ,{K_m}} \right\}, m \in {\cal M}$.

 Thus, problem~\eqref{POP} can be reformulated as:
 \begin{subequations}\label{POP-CVX}
\begin{align}
  &\mathop {\max }\limits_{{p_{m,k}}} \sum\nolimits_{m \in {\cal M}} {{R_{m,{K_m}}}},   \\
  &{s.t.} \ ~\eqref{CVX in POP},~\eqref{op d}. \label{POP-CVX-1}
\end{align}
\end{subequations}

It is noted that problem~\eqref{POP-CVX} is convex and can be solved efficiently by convex solvers such as CVX~\cite{grant2014cvx}.

The proposed joint active beamforming, passive beamforming and power allocation algorithm is summarized in \textbf{Algorithm 1}.
\begin{remark}\label{remark:algorithm}
The approximations of the active beamforming problem~\eqref{ABOP} and passive beamforming problem~\eqref{PBOP} are lower bounds for the original problem~\eqref{op}. Thus, the solution generated by \textbf{Algorithm 1} is sub-optimal. In each iteration of \textbf{Algorithm 1}, the objective value of problem~\eqref{op} is monotonically non-decreasing. Due to the transmit power constraint, the achievable sum rate sequence is upper bounded. Therefore, the proposed algorithm is guaranteed to converge.

  The complexities of solving the problems~\eqref{AB-SCA2},~\eqref{equivalent PB iterative} and~\eqref{POP-CVX} with the interior-point method at each iteration are $O\left( {{M^3}{{\left( {{N_T} + 2} \right)}^3} + K{M^2}{{\left( {{N_T} + 2} \right)}^2}} \right)$, $\left( {{{\left( {{L_{{\rm{IRS}}}} + 2M} \right)}^2}\left( {{L_{{\rm{IRS}}}} + 3M + K} \right)} \right)$ and $O\left( {{K^3}} \right)$, respectively.
\end{remark}

\begin{algorithm}
\caption{The Proposed Optimization Algorithm}
\label{The Proposed JAPP Algorithm}
\begin{algorithmic}[1]
  \STATE  Initialize a decoding order and feasible points $ {\theta _l^{\left( 0 \right)}}$, $ {{{p}}_{m,k}^{\left( 0 \right)}} $, $m \in \mathcal{M},k \in {\mathcal{K}_m},l \in {\mathcal{L}_{{\rm{IRS}}}}$. Let iteration index ${t} = 1$.
\REPEAT
  \STATE  update $ {{\textbf{w}}_m^{\left( t \right)}} $ by solving problem~\eqref{AB-SCA2} with $ {\theta _l^{\left( t-1 \right)}} $,  $ {{{p}}_{m,k}^{\left( {t - 1} \right)}} $;
  \STATE  update ${{\textbf{v}}^{\left( t \right)}}$ by solving problem~\eqref{equivalent PB iterative} with $ {\textbf{w}_m^{\left( t \right)}} $, $ {{{p}}_{m,k}^{\left( t-1 \right)}}$;
  \STATE  calculate $ {\theta _l^{\left( t \right)}} $ according to~\eqref{discrete theta};
  \STATE  update $ {p_{m,k}^{\left( t \right)}} $ by solving problem~\eqref{POP-CVX} with $ {\textbf{w}_m^{\left( t \right)}} $, $ {\theta _l^{\left( t \right)}} $;
  \STATE  ${t} = {t} + 1$;
\UNTIL {the objective value of problem~\eqref{op} converge.}
  \STATE   \textbf{Output}: optimal $ {{\textbf{w}}_m^{\left( t \right)}}, {\theta _l^{\left( t \right)}} , {{{p}}_{m,k}^{\left( t \right)}} $.
\end{algorithmic}
\end{algorithm}
\vspace{-0.5cm}
\section{Numerical Results}
Here, the performance of the proposed algorithm is evaluated through numerical simulations. Assume that the BS and IRS are located at coordinates (0 m, 0 m, 15 m) and (20 m, 20 m, 15 m), respectively. The mobile users are randomly and uniformly placed in a circle centered at (30 m, 30 m, 0 m) with radius 8 m. The complex channel gains $\mathfrak{p}_n^{\rm BI}$ and $\mathfrak{p}_{m,k,n}^{\rm IU} $ follow $\mathcal{CN}\left( {0,{{10}^{ - 0.1\mathcal{P}\left( {{d_{\rm TR}}} \right)}}} \right)$~\cite{cao2019intelligent,wang2019intelligent,xiu2020irs}, where $\mathcal{P}\left( {{d_{\rm TR}}} \right) = {\beta _1} + 10{\beta _2}{\log _{10}}\left( {{d_{\rm TR}}} \right) + {\beta _3}$, with $d_{\rm TR}$ as the distance between the transmitter and the receiver. The values of ${\beta _1}$, ${\beta _2}$ and ${\beta _3}$ at 28 GHz are set according to Table I in~\cite{akdeniz2014millimeter}. Let the number of propagation paths ${N_{\rm BI}} = {N_{\rm IU}} = 3$. Assume that the system bandwidth is 100 MHz, the noise power is ${\sigma ^2} =  -174~ \rm dBm$ and the QoS requirement is $R^{\rm min}_{m,k}=0.01~{\rm bit/s/Hz}$. The number of antennas at the BS is $N_{\rm T}=64$, the number of clusters is $M=3$ and each cluster contains ${K_m} = 3$ users.

\begin{figure}[t!]
 \setlength{\abovecaptionskip}{0cm}
 \setlength{\belowcaptionskip}{-0.3cm}   
 \centering
 \includegraphics[width=2.2in]{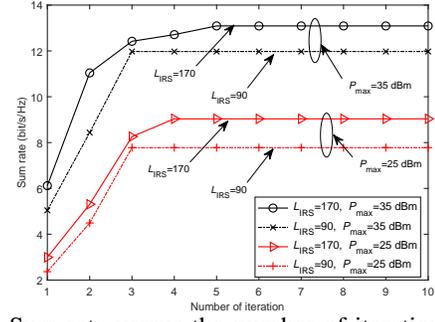}
  \caption{Sum rate versus the number of iterations, \emph{B}=5. }\label{sumR_vs_t}
 \end{figure}

The convergence of the proposed algorithm is depicted in Fig.~\ref{sumR_vs_t}. As can be observed, the proposed algorithm converges quickly within a small number of iterations under different settings of passive reflecting elements $L_{\rm IRS}$ and transmit power ${P_{\max }}$. This phenomenon is consistent with \textbf{Remark~\ref{remark:algorithm}}. In addition, the number of iterations for the convergence of the proposed algorithm increases with $L_{\rm IRS}$, because more variables have to be optimized.

To evaluate the performance of the proposed algorithm, we also consider the following benchmark algorithms: 1) \textbf{ZF based algorithm}: In this algorithm, the optimal decoding order is obtained by exhaustive search method, and the active beamforming is solved by zero-forcing (ZF) method. The passive beamforming and power allocation are solved by \textbf{{Algorithm 1}}. 2) \textbf{IRS-mmWave-OMA}: We also consider the IRS enhanced mmWave-OMA systems. There is also no direct links between BS and users. The joint passive beamforming and power allocation problem is solved by \textbf{{Algorithm 1}}, and the active beamforming is solved by the ZF method.

\begin{figure}[t!]
  \setlength{\abovecaptionskip}{0cm}
 \setlength{\belowcaptionskip}{-0.3cm}   
 \centering
 \includegraphics[width=2.2in]{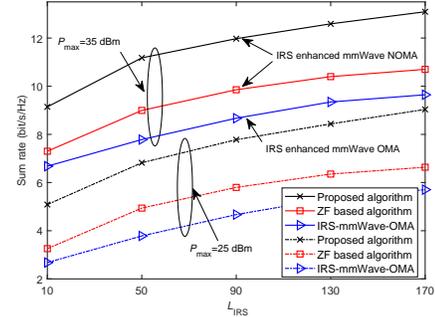}
  \caption{Sum rate versus $L_{\rm IRS}$, \emph{B}=5.}\label{sumR_vs_L}
 \end{figure}

Fig.~\ref{sumR_vs_L} plots  the system sum rate versus the number of passive reflecting elements $L_{\rm IRS}$. It is observed that the system sum rate achieved by all algorithms increases with $L_{\rm IRS}$. This indicates that more IRS passive reflecting elements can reflect more power of the signals received from the BS, which leads to more power gain. It is also seen that the proposed algorithm achieves better performance than the ZF-based algorithm. This is because the active beamforming is well optimized in our proposed algorithm, while the active beamforming in the ZF-based algorithm is obtained by the ZF method, which is not optimal for the considered problem. We further notice that the proposed algorithm outperforms the IRS-mmWave-OMA algorithm, since all NOMA users can be served simultaneously compared with the OMA system.

\begin{figure}[t!]
  \setlength{\abovecaptionskip}{0cm}
 \setlength{\belowcaptionskip}{-0.7cm}   
 \centering
 \includegraphics[width=2.2in]{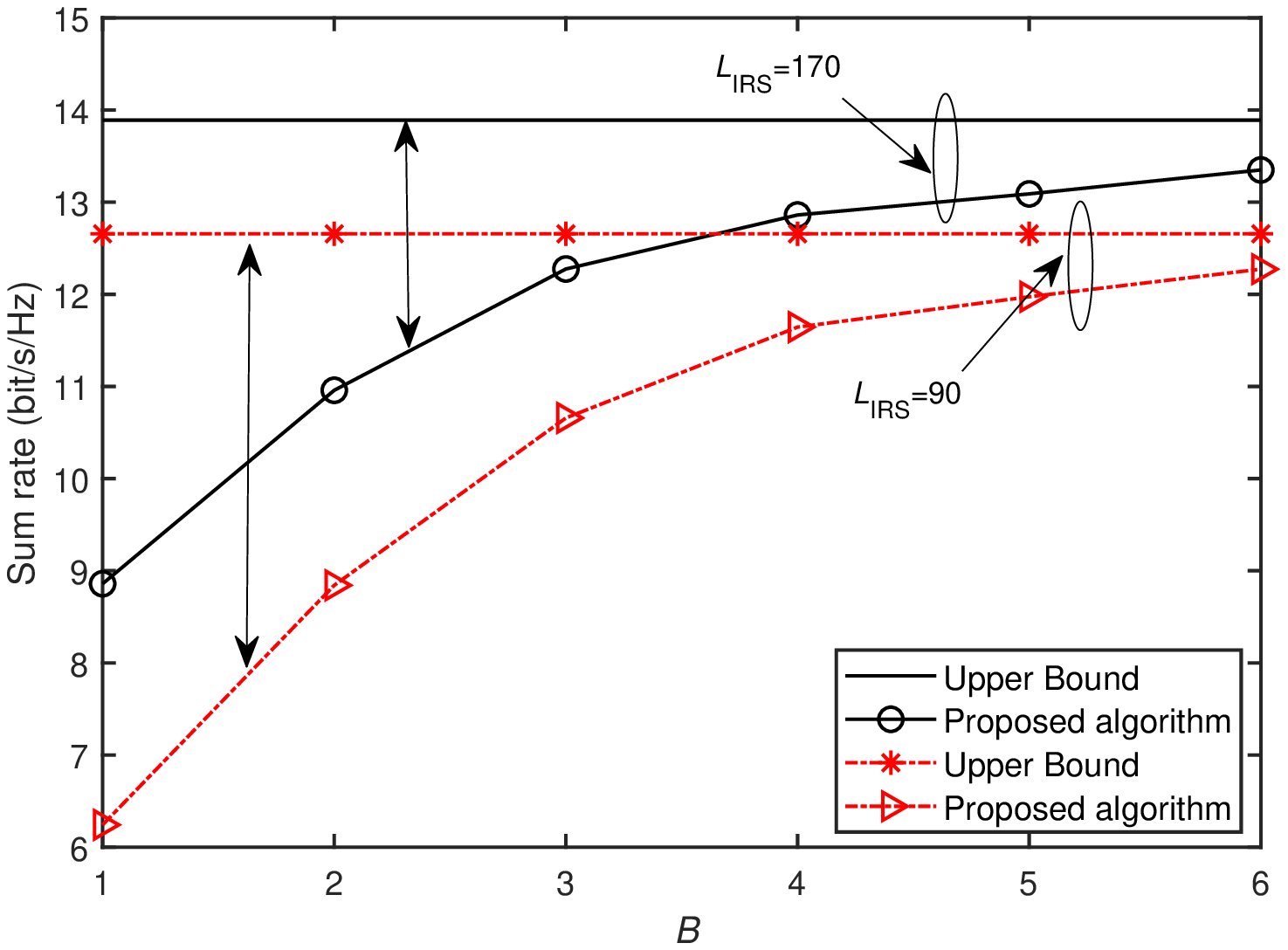}
  \caption{Sum rate versus $\emph{B}$, ${P_{\max }} = 35~{\rm dBm}$.}\label{sumR_vs_B}
 \end{figure}

Fig.~\ref{sumR_vs_B} depicts the impact of the phase shifts resolution bits $\emph{B}$ on the system sum rate. Note that 'Upper bound' scheme denotes \textbf{{Algorithm 1}} with continuous phase shifts, i.e., ${\theta _l} \in \left[ {0,2\pi } \right]$, $l \in {L_{{\rm{IRS}}}}$. It is observed that the system sum rate gap between the continuous and discrete phase shifts gradually decreases as the resolution bits \emph{B} increases. This is because a larger $\emph{B}$ value allows a better adjustment on the IRS phase shifts. This phenomenon is also confirmed by the insights in \textbf{Remark~\ref{remark:discrete}}. However, the implementation difficulty increases in practice with a higher number of resolution bits, leading to trade-off between the sum rate and number of resolution bits.
 \vspace{-0.6cm}
\section{Conclusion}
This paper proposed an IRS enhanced mmWave-NOMA system. The joint active beamforming, passive beamforming and power allocation optimization was investigated. The non-convex problem was decomposed into three sub-problems, which were solved by the alternative optimization and successive convex approximation. Simulation results showed that the proposed algorithm can improve the performance of the novel IRS assisted mmWave-NOMA system. Our results confirm that introducing IRS, the coverage of the assisted mmWave-NOMA system can be enhanced especially when there is no direct links between BS and users.
\vspace{-0.4cm}
\bibliographystyle{IEEEtran}
\bibliography{myref}

\end{document}